\documentclass[twoside,12pt]{article} 
\usepackage{indentfirst}
\usepackage{bm}    
\usepackage{amsmath}
\usepackage{amsthm}
\usepackage{amsfonts}
\usepackage{amssymb}
\usepackage{enumerate}

\begin{document}

\begin{center}
\LARGE\bf Deriving Some Quantum Optical Identities Using the General Ordering Theorem
\end{center}

\begin{center}
\rm Farid Sh\"ahandeh\footnote{Corresponding author. E-mail: shahandeh@ikiu.ac.ir} \ , \ Mohammad Reza Bazrafkan \ and \ Elahe Nahvifard 
\end{center}

\begin{center}
\begin{footnotesize} \sl
\textit{Physics Department, Faculty of Science, I. K. I. University, Qazvin, Iran}
\end{footnotesize}
\end{center}

\vspace*{2mm}

\begin{abstract}
Using the newly introduced general ordering theorem (GOT) by Sh\"ahandeh and Bazrafkan, we derive and generalize some quantum optical identities and give their applications.
\end{abstract}

\begin{center}
\begin{minipage}{15.5cm}
\begin{minipage}[t]{2.3cm}{\bf Keywords:}\end{minipage}
\begin{minipage}[t]{13.1cm}
$s$-ordered Expansion of Operators, General Ordering Theorem
\end{minipage}\par\vglue8pt
{\bf PACS: }
42.50.-p,~31.15.-p,~03.65.-w
\end{minipage}
\end{center}

\section{Introduction}
The $s$-ordering of operators has many applications in quantum optics such as constructing the phase space $s$-parameterized quasiprobabilities.~\cite{CG1,CG2} Perhaps the most well-known method of ordering of operators is that of FAN's IWSOP.~\cite{Fan1} However, in a recent work the authors have introduced the most general theorem regarding orderings of operators which will be referred to as the general ordering theorem or just GOT.~\cite{SB} In this regard, one may $t$-order any multiplicative sequence of $s_j$-ordered functions with $j \in \left\{ {2,3, \ldots ,k} \right\}$ as
	\begin{eqnarray}
	\label{sjt}
	{\left\{ {\hat F\left( {a^\dag,a} \right)} \right\}_{{s_2}}}{\left\{ {\hat G\left( {a^\dag,a} \right)}\right\}_{{s_3}}}\cdots 	{\left\{ {\hat H\left( {a^\dag,a} \right)} \right\}_{{s_k}}} = \nonumber \\ 								\sum\limits_{{\text{$i$-pair $\left({{u_l},t} \right)$-contractions}}}{{\left\{ {{\left( {\hat F\hat G \cdots \hat H} \right)}_i^{\mathbf{u}}\left( {a^\dag, a} \right)} \right\}}_t}~,
	\end{eqnarray}
in which $l \in \left\{ {1,2, \ldots ,i} \right\}$, ${\mathbf{u}} \equiv \left( {{{u}_1},{{u}_2} \ldots ,{{u}_i}} \right)$ with $u_l \in \left\{ {{-1,1,s_j}} \right\}$, and ${\left( {\hat F\hat G \cdots \hat H} \right)_i^{{\mathbf{u}}}}$ is the $i$-pair $\left (u_l,t\right )$-contracted form of the multiplicative sequence $\hat F\hat G \cdots \hat H$.

This statement has an obvious meaning by virtue of the definition of relative order of operators. Strictly speaking, considering a multiplicative sequence of $s$-ordered monomials, each pair of operators $a^\dag$ and $a$ have a relative order depending on their relative positions:
	\begin{enumerate}
	\item If $a^\dag$ stands on the left of $a$ lying in two different $s$-ordering symbols, they are relatively normally ordered, specified by the parameter $s=1$.
	\item If $a$ stands on the left of $a^\dag$ lying in two different $s$-ordering symbols, they are relatively anti-normally ordered, specified by the parameter $s=-1$.
	\item and, if $a^\dag$ and $a$ both lie in the same ${\left\{  \cdots  \right\}_s}$ symbol, they are relatively $s$-ordered.
	\end{enumerate}
In this regard, for two operators $a^\dag$ and $a$ with $\left [a, a^\dag \right ]=1$ a $\left ({s,t} \right )$-contraction is defined to be the replacement of a pair of relatively $s$-ordered operators by
	\begin{equation} \label{cont}
	{\left\{ {a^\dag a} \right\}_s} - {\left\{{a^\dag a}\right\}_t} = \frac{{t - s}}{2}~,
	\end{equation}
where the right-hand-side is a direct result of the $s$-ordering definition.

The relation~\eqref{sjt} provides the combinatorial structure underlying the ordering procedure. In fact, this is a generalization of the Wick's idea of normal ordering. In this way, one may $s$-order any given function granted that he could count the number of different contractions.

In the present paper, we derive some useful quantum optical identities by virtue of the above theorem.

\section{Factorial Identities}

We may simply derive Glauber's formula~\cite{Glauber}
	\begin{equation}
	 {\lambda ^{{a^\dag }a}}=:{e^{\left( {\lambda  - 1} \right){a^\dag }a}}:~.
	\end{equation}
To this end, we write
	\begin{eqnarray*}
	\left( {{a^\dag }a} \right):{\left( {{a^\dag }a} \right)^{n - 1}}: &=& \sum\limits_{i = 0}^{\min \left\{ {1,n-1} \right\}} {\left( {\begin{array}{*{20}{c}}
  1 \\ 
  i 
\end{array}} \right)\left( {\begin{array}{*{20}{c}}
  {n - 1} \\ 
  i 
\end{array}} \right)i!{{\left( 1 \right)}^i}:{{\left( {{a^\dag }a} \right)}^{n - i}}:}  \nonumber \\
&=& :{\left( {{a^\dag }a} \right)^n}: + \left( {n - 1} \right):{\left( {{a^\dag }a} \right)^{n - 1}}:~,
	\end{eqnarray*}
in which the coefficient $\left( {\begin{array}{*{20}{c}}
  1 \\ 
  i 
\end{array}} \right)\left( {\begin{array}{*{20}{c}}
  n \\ 
  i 
\end{array}} \right)i!$ is just the number of possible contractions between the left $a$ with the inside $a^\dag$s. This, indeed, gives
	\begin{equation*}
	:{\left( {{a^\dag }a} \right)^n}: = \left( {{a^\dag }a - n + 1} \right):{\left( {{a^\dag }a} \right)^{n - 1}}:~,
	\end{equation*}
and then,
	\begin{equation*}
	:{\left( {{a^\dag }a} \right)^n}: = \frac{{\left( {{a^\dag }a} \right)!}}{{\left( {{a^\dag }a - n} \right)!}}~.
	\end{equation*}
Now, we have
	\begin{eqnarray} \label{eL}
	:{e^{\left( {\lambda  - 1} \right){a^\dag }a}}: &=& \sum\limits_{n = 0}^\infty  {\left( {\begin{array}{*{20}{c}}
  {{a^\dag }a} \\ 
  n 
\end{array}} \right){{\left( {\lambda  - 1} \right)}^n}}  \hfill \nonumber \\
&=& {\lambda ^{{a^\dag }a}}~,
	\end{eqnarray}
in which we have used the binomial formula. In the same manner, one may write
	\begin{eqnarray*}
	\left( {a{a^\dag }} \right) \vdots {\left( {{a^\dag }a} \right)^{n - 1}} \vdots  &=& \sum\limits_{i = 0}^n {\left( {\begin{array}{*{20}{c}}
  1 \\ 
  i 
\end{array}} \right)\left( {\begin{array}{*{20}{c}}
  {n - 1} \\ 
  i 
\end{array}} \right)i!{{\left( { - 1} \right)}^i} \vdots {{\left( {{a^\dag }a} \right)}^{n - i}} \vdots } \nonumber \\
&=&  \vdots {\left( {{a^\dag }a} \right)^n} \vdots  - \left( {n - 1} \right) \vdots {\left( {{a^\dag }a} \right)^{n - 1}} \vdots~,
	\end{eqnarray*}
which leads to
	\begin{equation} \label{antiLambda}
	\vdots {\left( {{a^\dag }a} \right)^n} \vdots = \frac{{\left( {{a^\dag }a + n} \right)!}}{{\left( {{a^\dag }a} \right)!}}~.
	\end{equation}
Thus, we may use the negative-binomial relation,
	\begin{equation*}
	\sum\limits_{n = 0}^\infty  {\left( {\begin{array}{*{20}{c}}
  {m + n} \\ 
  n 
\end{array}} \right){{\left( { - x} \right)}^n}}  = {\left( {x + 1} \right)^{ - \left( {m + 1} \right)}}~,
	\end{equation*}
together with Eq.~\eqref{antiLambda} to write
	\begin{equation}
	\lambda  \vdots {e^{\left( {1 - \lambda } \right){a^\dag }a}} \vdots  = {\lambda ^{ - {a^\dag }a}}~.
	\end{equation}

\section{Generalized Derivative Lemmas}

In this section, we use the GOT to generalize some derivative formulas. We begin with the product ${a^n}\hat F\left( {{a^\dag },a} \right)$. Consider the case for which we know the $s$-ordered form of $\hat F\left( {{a^\dag },a}\right)$ to be ${\hat F^{\left( s \right)}}\left( {{a^\dag },a} \right)$. Then,
	\begin{eqnarray*}
	{a^n}{\hat F^{\left( s \right)}}\left( {{a^\dag },a} \right) &=& {a^n}\sum\limits_{k,l} {{f_{kl}^s}\left\{{a^{\dag k}}{a^l}\right\}_s} \nonumber \\
&=& {\left\{ {\sum\limits_{k,l} {{f_{kl}^s}\sum\limits_{i = 0}^{\min \left\{ {k,n} \right\}}  {\left( {\begin{array}{*{20}{c}}
  k \\ 
  i 
\end{array}} \right)\left( {\begin{array}{*{20}{c}}
  n \\ 
  i 
\end{array}} \right)i!{{\left( {\frac{{s + 1}}{2}} \right)}^i}{a^{\dag k - i}}{a^{n + l - i}}} } } \right\}_s}~.
	\end{eqnarray*}
Now, we may substitute the coefficient $\left( {\begin{array}{*{20}{c}}
  k \\ 
  i 
\end{array}} \right)i!$ by the formal differentiation $\partial _{{a^\dag }}^i$ to write
	\begin{equation} \label{anF}
	{a^n}{{\hat F}^{\left( s \right)}}\left( {{a^\dag },a} \right) = {\left\{ {{{\left[ {a + \left( {\frac{{s + 1}}{2}} \right)\frac{\partial }{{\partial {a^\dag }}}} \right]}^n}{F^{\left( s \right)}}\left( {{a^\dag },a} \right)} \right\}_s}~.
	\end{equation}
This is a generalization of the well-known derivative lemma,~\cite{Mandel} ($s$=1)
	\begin{equation}
	{a^n}{{\hat F}^{\left( N \right)}}\left( {{a^\dag },a} \right) = :{\left( {a + \frac{\partial }{{\partial {a^\dag }}}} \right)^n}{F^{\left( N \right)}}\left( {{a^\dag },a} \right):~,
	\end{equation}
with the trivial case of $s=-1$ as
	\begin{equation} \label{antianF}
	{a^n}{{\hat F}^{\left( A \right)}}\left( {{a^\dag },a} \right) =  \vdots {a^n}{F^{\left( A \right)}}\left( {{a^\dag },a} \right) \vdots~.
	\end{equation}

In a similar way, we may write
	\begin{eqnarray} \label{Fan}
  {{\hat F}^{\left( s \right)}}\left( {{a^\dag },a} \right){a^n} &=& \sum\limits_{k,l} {f_{kl}^s\left\{{a^k}{a^{\dag l}}\right\}_s} {a^n} \nonumber \\ 
&=& {\left\{ {\sum\limits_{k,l} {f_{kl}^s\sum\limits_{i = 0}^{\min \left\{ {l,n} \right\}}  {\left( {\begin{array}{*{20}{c}}
  n \\ 
  i 
\end{array}} \right)\left( {\begin{array}{*{20}{c}}
  l \\ 
  i 
\end{array}} \right)i!{{\left( {\frac{{s - 1}}{2}} \right)}^i}{a^{\dag l - i}}{a^{n + k - i}}} } } \right\}_s} \nonumber \\ 
&=& {\left\{ {{{\left[ {a + \left( {\frac{{s - 1}}{2}} \right)\frac{\partial }{{\partial {a^\dag }}}} \right]}^n}{F^{\left( s \right)}}\left( {{a^\dag },a} \right)} \right\}_s}~,
	\end{eqnarray}
which in the special case of $s=-1$ gives the formula
	\begin{equation}
	{{\hat F}^{\left( A \right)}}\left( {{a^\dag },a} \right){a^n} =  \vdots {\left( {a - \frac{\partial }{{\partial {a^\dag }}}} \right)^n}{F^{\left( A \right)}}\left( {{a^\dag },a} \right) \vdots~.
	\end{equation}
Alternatively, one may examine multiplication by $a^{\dag n}$. This leads to
	\begin{eqnarray}
	{a^{\dag n}}{{\hat F}^{\left( s \right)}}\left( {{a^\dag },a} \right) = {\left\{ {{{\left[ {{a^\dag } + \left( {\frac{{s - 1}}{2}} \right)\frac{\partial }{{\partial a}}} \right]}^n}{{\hat F}^{\left( s \right)}}\left( {{a^\dag },a} \right)} \right\}_s} \label{aDnF}~, \\
	{{\hat F}^{\left( s \right)}}\left( {{a^\dag },a} \right){a^{\dag n}} = {\left\{ {{{\left[ {{a^\dag } + \left( {\frac{{s + 1}}{2}} \right)\frac{\partial }{{\partial a}}} \right]}^n}{{\hat F}^{\left( s \right)}}\left( {{a^\dag },a} \right)} \right\}_s} \label{FaDn}~,
	\end{eqnarray}
with the special cases
	\begin{eqnarray}
  {a^{\dag n}}{{\hat F}^{\left( A \right)}}\left( {{a^\dag },a} \right) &=& \vdots {\left( {{a^\dag } - \frac{\partial }{{\partial a}}} \right)^n}{{\hat F}^{\left( A \right)}}\left( {{a^\dag },a} \right) \vdots~,  \\ 
  {{\hat F}^{\left( N \right)}}\left( {{a^\dag },a} \right){a^{\dag n}} &=& :{\left( {{a^\dag } + \frac{\partial }{{\partial a}}} \right)^n}{{\hat F}^{\left( N \right)}}\left( {{a^\dag },a} \right):~.
	\end{eqnarray}
These are solely new identities regarding transformations of ordered operators. For example, one may use Eqs.~\eqref{anF} and~\eqref{aDnF} to obtain
	\begin{eqnarray}
	{e^{\lambda a}}{{\hat F}^{\left( s \right)}}\left( {{a^\dag },a} \right) = {\left\{ {{e^{\lambda a}}{{\hat F}^{\left( s \right)}}\left( {{a^\dag } + {\tau _ + }\lambda ,a} \right)} \right\}_s},\qquad {\tau _ + } \equiv \left( {\frac{{s + 1}}{2}} \right)~, \\
{e^{\lambda {a^\dag }}}{{\hat F}^{\left( s \right)}}\left( {{a^\dag },a} \right) = {\left\{ {{e^{\lambda {a^\dag }}}{{\hat F}^{\left( s \right)}}\left( {{a^\dag } + {\tau _ - }\lambda ,a} \right)} \right\}_s},\qquad {\tau _ - } \equiv \left( {\frac{{s - 1}}{2}} \right)~.
	\end{eqnarray}

\section{FAN's Hermite Polynomials}

We may also use the GOT to simply write down FAN's ordering formulas regarding Hermite polynomials as well.~\cite{Fan2} In this regard, we have
	\begin{equation*}
	{a^{\dag n}}{a^m} = {\left\{ {\sum\limits_{i = 0}^{\min \left\{ {n,m} \right\}}  {\left( {\begin{array}{*{20}{c}}
  n \\ 
  i 
\end{array}} \right)\left( {\begin{array}{*{20}{c}}
  m \\ 
  i 
\end{array}} \right)i!{{\left( {\frac{{s - 1}}{2}} \right)}^i}{a^{\dag n - i}}{a^{m - i}}} } \right\}_s}~,
	\end{equation*}
which using the definition of two-variable Hermite polynomials,~\cite{Erdelyi}
	\begin{equation} \label{tvHp}
	{H_{n,m}}\left( {x,y} \right) = \sum\limits_{i = 0}^{\min \left\{ {n,m} \right\}}  {\left( {\begin{array}{*{20}{c}}
  n \\ 
  i 
\end{array}} \right)\left( {\begin{array}{*{20}{c}}
  m \\ 
  i 
\end{array}} \right)i!{{\left( { - 1} \right)}^i}{x^{n - i}}{y^{m - i}}}~,
	\end{equation}
gives
	\begin{equation} \label{FANform1}
	{a^{\dag n}}{a^m} = {\left\{ {{{\left( {\frac{{1 - s}}{2}} \right)}^{\frac{{m + n}}{2}}}{H_{n,m}}\left( {\sqrt {\frac{2}{{1 - s}}} {a^\dag },\sqrt {\frac{2}{{1 - s}}} a } \right)} \right\}_s}~.
	\end{equation}
In a similar way, one may write the general transformation formula
	\begin{eqnarray} \label{FANform2}
	{\left\{ {{a^{\dag n}}{a^m}} \right\}_s} &=& {\left\{ {\sum\limits_{i = 0}^{\min \left\{ {n,m} \right\}}  {\left( {\begin{array}{*{20}{c}}
  n \\ 
  i 
\end{array}} \right)\left( {\begin{array}{*{20}{c}}
  m \\ 
  i 
\end{array}} \right)i!{{\left( {\frac{{t - s}}{2}} \right)}^i}{a^{\dag n - i}}{a^{m - i}}} } \right\}_t} \nonumber \\
&=& {\left\{ {{{\left( {\frac{{s - t}}{2}} \right)}^{\frac{{m + n}}{2}}}{H_{m,n}}\left( {\sqrt {\frac{2}{{s - t}}} a,\sqrt {\frac{2}{{s - t}}} {a^\dag }} \right)} \right\}_t}~.
	\end{eqnarray}
 
One could, indeed, use the incomplete two-variable Hermite polynomials,~\cite{Dattoli}
	\begin{equation}
	{h_{m,n}}\left( {x,y|\tau } \right) = \sum\limits_{i = 0}^{\min \left\{ {n,m} \right\}}  {\left( {\begin{array}{*{20}{c}}
  m \\ 
  i 
\end{array}} \right)\left( {\begin{array}{*{20}{c}}
  n \\ 
  i 
\end{array}} \right)i!{\tau ^i}{x^{m - i}}{y^{n - i}}}~,
	\end{equation}
with the generating function
	\begin{equation}\label{IHgen}
	\sum\limits_{m,n = 0}^\infty  {\frac{{{\lambda ^m}{\mu ^n}}}{{m!n!}}{h_{m,n}}\left( {x,y|\tau } \right)}  = {e^{\lambda x + \mu y + \tau \lambda \mu }}~,
	\end{equation}
to write Eqs.~\eqref{FANform1} and~\eqref{FANform2} as
	\begin{eqnarray}
	{a^{\dag n}}{a^m} = {\left\{ {{h_{n,m}}\left( {{a^\dag },a|{\tau _ - }} \right)} \right\}_s}~, \\
\left\{ {{a^{\dag n}}{a^m}} \right\}_s = {\left\{ {{h_{n,m}}\left( {{a^\dag },a|\tau_{st}} \right)} \right\}_t},\qquad \tau _{st} \equiv \frac{{t - s}}{2}~. \label{aDnam_s}
	\end{eqnarray}

\section{Application of the Results}

Using the definition of two-variable Hermite polynomials Eq.~\eqref{tvHp} we have
	\begin{equation}
	{h_{m,n}}\left( {x,y|\kappa } \right) = {\left( { - i\sqrt \kappa  } \right)^{m + n}}{H_{m,n}}\left( {i\frac{x}{{\sqrt \kappa}},i\frac{y}{{\sqrt \kappa  }}} \right)~.
	\end{equation}
As a special case, one has 
	\begin{equation}
	{h_{m,n}}\left( {x,y| - 1} \right) = {H_{m,n}}\left( {x,y} \right)~.
	\end{equation}

The relation between two-variable Hermite polynomials and generalized Laguerre polynomials $L_n^{m}\left( {x} \right)$,~\cite{Erdelyi}
	\begin{equation}
	{H_{m,n}}\left( {x,y} \right) = {\left( { - 1} \right)^n}n!{x^{m - n}}L_n^{m - n}\left( {xy} \right) = {\left( { - 1} \right)^m}m!{y^{n - m}}L_m^{n - m}\left( {xy} \right)~,
	\end{equation}
implies
	\begin{equation}
	{h_{m,n}}\left( {x,y|\kappa } \right) = {\kappa ^n}n!{x^{m - n}}L_n^{m - n}\left( { - \frac{{xy}}{\kappa }} \right) = {\kappa ^m}m!{y^{n - m}}L_n^{n - m}\left( { - \frac{{xy}}{\kappa }} \right)~.
	\end{equation}
This gives the special case of $m=n$ as
	\begin{equation} \label{specialm}
	{h_{n,n}}\left( {x,y|\kappa } \right) = {\kappa ^n}n!{L_n}\left( { - \frac{{xy}}{\kappa }} \right)~,
	\end{equation}
in which ${L_n}\left( {x} \right)$ are just the usual Laguerre polynomials.

Using Eqs.~\eqref{aDnam_s},~\eqref{specialm} and the generating function of the usual Laguerre polynomials~\cite{Bayin} we may write
	\begin{eqnarray} \label{eLaDa_s_t}
	{\left\{ {{e^{\lambda {a^\dag }a}}} \right\}_s} &=& {\left\{ {\sum\limits_{n = 0}^\infty  {\frac{{{\lambda ^n}}}{{n!}}{h_{n,n}}\left( {{a^\dag },a|{\tau _{st}}} \right)} } \right\}_t} \nonumber \\ 
   &=& \frac{1}{{1 - \lambda {\tau _{st}}}}{\left\{ {{e^{ \frac{\lambda }{{1 - \lambda {\tau _{st}}}}{a^\dag }a}}} \right\}_t}~.
	\end{eqnarray}

In addition, one may multiply both sides of Eq.~\eqref{eLaDa_s_t} on the left by $a^n$ and employ the relation~\eqref{anF} to get to
	\begin{eqnarray}
	{a^n}{\left\{ {{e^{\lambda {a^\dag }a}}} \right\}_s} = {\left[ {\frac{{1 - \lambda \left( {{\tau _{st}} + {\tau '_ + }} \right)}}{{{{\left( {1 - \lambda {\tau _{st}}} \right)}^2}}}} \right]^n}{\left\{ {{a^n}{e^{ \frac{\lambda }{{1 - \lambda {\tau _{st}}}}{a^\dag }a}}} \right\}_t} \label{aneL_s}~,\qquad
{\tau '_ +} \equiv \frac{{t + 1}}{2}~.
	\end{eqnarray}
Similarly, multiplying both sides of Eq.~\eqref{aneL_s} on the right by $a^{\dag m}$, after using~\eqref{FaDn}, leads to
	\begin{equation*}
	{a^n}{\left\{ {{e^{\lambda {a^\dag }a}}} \right\}_s}{a^{\dag m}} = {\left[ {\frac{{1 - \lambda \left( {{\tau _{st}} + {\tau '_ + }} \right)}}{{{{\left( {1 - \lambda {\tau _{st}}} \right)}^2}}}} \right]^n}{\left\{ {{{\left( {{a^\dag } + {\tau '_ + }{\partial _a}} \right)}^m}{a^n}{e^{\frac{\lambda }{{1 - \lambda {\tau _{st}}}}{a^\dag }a}}} \right\}_t}~,
	\end{equation*}
which by the substitution ${a^n} \to {\left( {\frac{{1 - \lambda {\tau _{st}}}}{\lambda }} \right)^n}\partial _{{a^\dag }}^n$ and using the general Leibniz rule of differentiation gives
	\begin{equation}
	{a^n}{\left\{ {{e^{\lambda {a^\dag }a}}} \right\}_s}{a^{\dag m}} = {\left[ {\frac{{1 - \lambda \left( {{\tau _{st}} + {\tau '_ + }} \right)}}{{\lambda \left( {1 - \lambda {\tau _{st}}} \right)}}} \right]^{n + m}}{\lambda ^m}{\left\{ {{h_{m,n}}\left( {\left. {{a^\dag },\frac{\lambda }{{1 - \lambda {\tau _{st}}}}{a^\dag }} \right|1} \right){e^{\frac{\lambda }{{1 - \lambda {\tau _{st}}}}{a^\dag }a}}} \right\}_t}~.
	\end{equation}
This is the most general ordered product rule of this kind.

\section{Conclusion}

In summary, we have obtained a simple way of $s$-ordering of operators. This is specially suitable for evaluating product of ordered operators, as well as ordering transformations. Indeed, we have re-derived and generalized some elementary quantum optical identities in a simple fashion. We have shown that some of the previous ordering identities have simple combinatorial meanings. We have also represented the extent of our method through a simple example.

\end{document}